\documentstyle[preprint,aps,psfig]{revtex}

\begin{document}

\title{Viscosity Dependence of the Folding Rates of Proteins}
\author{D.K.Klimov and D.Thirumalai}

\address{Institute for Physical Science and Technology and 
Department of Chemistry and Biochemistry\\
University of Maryland, College Park, Maryland 20742}

\maketitle

\begin{abstract}
The viscosity (\(\eta \)) dependence of the folding rates for four
sequences (the native state of three sequences is a \(\beta \)-sheet,
while the fourth forms a \(\alpha \)-helix) is
calculated  for off-lattice models of proteins.    
Assuming that the dynamics is given by
the Langevin equation we show that the folding rates 
increase linearly at low
viscosities 
\(\eta \), decrease as \(1/\eta \) at large \(\eta \) and have a maximum at
intermediate values. The Kramers theory of barrier crossing provides
a {\em quantitative} fit of the numerical results. By mapping the
simulation results to real proteins we estimate that for optimized
sequences the time scale
for forming a four turn \(\alpha
\)-helix topology is about \(500\,ns\), whereas for \(\beta \)-sheet it
is about \(10\,\mu s\). 
\end{abstract}
\newpage

Based on several theoretical studies of minimal models of proteins a
novel conceptual framework for understanding the folding kinetics of
proteins \cite{Wol,DillChan,Bryn95,Dill,Thirum95} and, in general,  
biomolecules \cite{ThirumWood} 
has  recently
emerged. The basis of this new framework lies in the observation that
the polymeric nature of proteins together with the presence of
conflicting energies (arising from the differences in the structural
preferences  of hydrophobic and
hydrophilic residues)  lead to topological frustration - a situation
where  structures, which are favorable  on relatively short length
scales, are in conflict with the global free energy minimum, namely,
the native state of the protein.  
Due to topological frustration the underlying
free energy landscape has, besides the dominant native basin of
attraction (NBA), several competing  basins of attraction (CBA). 
Theoretical  studies of folding kinetics in such a complex energy
landscape suggest 
that, in general, the folding of biomolecules takes place by 
multiple pathways rather than by a hierarchical organization.  Recent
experiments lend support to the "new view" 
\cite{Wol,DillChan,Bryn95,Dill,Thirum95,ThirumWood} of the folding of
biomolecules \cite{Rad}.

One of the important  theoretical predictions is that the mechanisms of
protein folding can be varied depending not only on the intrinsic 
sequence properties but also on
external conditions (such as pH, salt concentration, viscosity etc.) 
\cite{Bryn95,Thirum95,ThirumWood}. 
The theoretical studies 
to date have focused on the temperature dependence
of folding rates using minimal models of proteins. 
The purpose of
this paper is to examine the dependence of the rates of protein
folding on viscosity, \(\eta \), (or equivalently the friction
coefficient \(\zeta \)) and to provide a picture of the folding
process in terms of the free energy landscape classified using 
NBA and CBA. 
Although there are a  few experimental studies that have probed the  
dependence of the folding
rates on viscosity \cite{Wald} they have not been systematic enough to
reveal the underlying folding mechanisms.

We use continuum minimal model
representation of the polypeptide chain and Langevin  dynamics 
to compute  
the folding rates as a function of viscosity. 
The major results of this study, which were obtained by examining four
sequences each with either a \(\beta \)-sheet or \(\alpha\)-helix as
the native state, are: \\
(a) The folding rate \(k_{F}\) for the formation of a \(\beta \)-sheet
or a \(\alpha \)-helix increases linearly with 
\(\eta \) at low
viscosities reaching a maximum at moderate values of \(\eta \) and
starts to decrease as \(1/\eta \) at higher viscosities. \\
(b) By assuming that the typical free energy barrier to folding scales as
\(\sqrt{M} k_{B}T_{s}\) \cite{Thirum95} 
(\(T_{s}\) is the simulation temperature and \(M\) is the number of
beads in the sequence) 
we find that the Kramers expression for 
barrier crossing \cite{Hanggi} 
(with the frequencies at the bottom and at the barrier height of an
appropriate reaction coordinate as
adjustable parameters) gives a {\em quantitative} fit of the
simulation results. This implies that, at least for small  proteins
with simple native state topology, a low dimensional (or even one)
reaction coordinate can adequately describe the folding process
\cite{Socci96}. \\
(c) For fast folding sequences, i.e. those that essentially
display two-state folding kinetics, 
the fraction of molecules that reaches the native state rapidly
without being trapped in CBA, 
namely, the partition factor \(\Phi \) is close to unity and is 
{\em independent} of 
viscosity.  For slow and moderate folders \(\Phi \) depends on \(\eta
\) implying that viscosity can be used to alter the mechanisms of
protein folding.

The polypeptide chain is modeled  
by a sequence of \(M\) connected beads, each of them corresponding to,
perhaps, a blob of actual \(\alpha \)-carbons \cite{Veit}. 
The chain conformation is
determined by the vectors \(\{\vec{r}_{i}\}\),
\(i=1,2...M\). 
Although real polypeptide
sequences are made from twenty amino acids, it has been shown that
a three letter code sequences (i.e., sequences of residues of 
three types) can faithfully mimic certain properties of real
proteins \cite{Veit}. Accordingly, we assume that  protein 
sequence is made of 
hydrophobic (\(B\)), hydrophilic (\(L\)), and neutral (\(N\)) residues. 
In these models a sequence is specified by the precise way in which
\(B\), \(L\), and \(N\) beads are connected together.

Following our earlier work \cite{Veit} 
the energy of a conformation is taken to
be the sum of bond-stretch potential, bond-angle potential,
potential associated with the dihedral angle degrees of freedom  
and non-bonded
potential, which is responsible for tertiary interactions. 
The details of the potentials and the compositions of the three
sequences labeled E, G, and I  
with a \(\beta\)-sheet as the native state are given elsewhere 
\cite{Veit}. The potential energy function and sequence composition 
for sequence H with the
\(\alpha \)-helix  as the native conformation are the same
as in the previous study \cite{Guo4Helix}. The parameters in the
dihedral angle potential 
\( V(\phi)\,=\,A_{\phi} (1-cos\phi) + B_{\phi}(1+cos3\phi) + C_{\phi}
(1-sin\phi) \) are taken to be \(A_{\phi}=1.0\,\epsilon_{h}\),
\(B_{\phi}=1.6\,\epsilon_{h}\), \(C_{\phi}=2.0\,\epsilon_{h}\), 
respectively, where the parameter 
\(\epsilon_{h} \approx (1\,-\,2)\,kcal/mol\) is the average
strength of the hydrophobic interaction. 
These parameters differ from the ones used
earlier \cite{Guo4Helix}. 
The native conformations for sequences I and H, which are  
determined by the methods described in 
\cite{Veit}, are displayed in Fig. (1). 

The choice of sequences was dictated by the following considerations. 
It has been shown \cite{Veit} that the kinetic accessibility and 
the associated thermodynamic stability of the native
conformation for minimal protein models correlate extremely well
with \(\sigma = (T_{\theta } - T_{F})/T_{\theta }\) (\(0\lesssim
\sigma \lesssim 1\)), where the two
characteristic temperatures intrinsic to the sequence, \(T_{\theta }\)
and \( T_{F}\), are the collapse and the folding transition temperatures,
respectively. Sequences with relatively small values of
\(\sigma \) reach the native conformation very rapidly without being
trapped in any detectable intermediates, whereas  those sequences with
large \(\sigma \) have several CBA that act as kinetic traps \cite{Veit}. 
Two of the sequences, G and I, have relatively small values of \(\sigma
\) \cite{Veit} (0.20 and 0.14, respectively) and hence are fast folders, 
which implies that in excess of 90\% of the
molecules reach the native conformation  on the time scale in which
collapse and formation of the  native state are almost synchronous 
\cite{Thirum95}. 
Sequence E is a  
moderate folder, while H is a slow folder (with \(\sigma = 0.39\) and
\(0.75\), respectively). 
Since these four sequences involve the two common structural motifs in
proteins and span a range of \(\sigma \) meaningful conclusions
regarding the viscosity dependence of small proteins can be drawn. 

We assume that Langevin equation provides an adequate description of
the polypeptide chain dynamics. 
Since our
goal is to study the dependence of the folding rate as a function of
viscosity over a wide range,  we are forced to use different algorithms
depending on the precise value of the viscosity \(\eta \) or the friction
coefficient \(\zeta \)(\(=6 \pi \eta  a\)). In the low \(\zeta \)
limit, corresponding to the energy diffusion regime, the 
inertial terms are important and we use the noisy molecular
dynamics \cite{Veit}. 
At higher values of \(\zeta \) we
use the Brownian dynamics algorithm of Ermak and
McCammon \cite{Ermak}. We have verified that both the 
algorithms give
{\em identical} results for folding rates in the intermediate range
of \(\zeta \). 
For the sake of consistency we measure time in 
units of \(\tau _{L} = (ma^{2}/\epsilon _{h})^{1/2}\), where \(m\) is
the mass of a bead and \(a\) is the bond length between successive
beads. 

The external conditions in the simulations are \(\zeta \) and the
temperature \(T\). Since we focus here on the variation of the
folding rate with \(\zeta \) it is desirable to choose sequence
dependent simulation temperatures \(T_{s}\) so that the extent of the
native conformation (given by the structural overlap function \(\chi
\), which measures the similarity of a given conformation to the native
state \cite{Veit}) is the same for all the sequences. 
The simulation temperature, \(T_{s}\), is chosen so
that (i) \(T_{s} < T_{F}\) and (ii) \(<\chi (T_{s})> = \alpha \) be
the same for all sequences with a given native topology. 
The condition \(T_{s} < T_{F}\) ensures that the native conformation has
the largest occupation probability, while the second condition (ii)
allows us to subject the sequences to similar folding conditions. 
With the assumption that \(\alpha = 0.26\),
\(T_{s}\) (measured in the units of \(\epsilon _{h}\)) turns 
out to be 0.29, 0.37, and 0.41 for sequences E, I, and G, respectively
\cite{Veit}. For sequence H we took \(\alpha \) to be 0.32, 
and the resulting  \(T_{s}\) is 0.24 (details to be published
elsewhere).

The folding rate  is calculated as 
\begin{equation}
k_{F}=\frac{1}{N_{max}} \sum _{i=1}^{N_{max}} \frac{1}{\tau _{1i}}
\end{equation}
where \(\tau _{1i}\) is the first passage time (the first time a given 
trajectory reaches the native conformation) for the trajectory
\(i\) and \(N_{max}\) is the maximum number of trajectories used in
the simulations. The value of \(N_{max}\) ranges from 200-600, 
depending on \(\zeta \) and the sequence, which gives well converged results for \(k_{F}\). 
In Fig. (2) we plot the ratio \(k_{F}/k_{TST}\) as a function of \(\zeta
\), where \(k_{TST}\) is the transition state estimate of the folding
rate. The calculation of \(k_{TST}\) is described in the caption to
Fig. (2). 
The top, middle, and lower panels correspond to sequences G, E, and H,
respectively. 
It is clear that the folding rate increases (roughly
linearly) at low \(\zeta \)  and decreases as \(\zeta ^{-1}\) at higher
viscosity. There is a maximum at moderate values of \(\zeta \). 

The remarkable similarity of the dependence of the 
rate of folding to the predictions of Kramers
theory for barrier crossing \cite{Hanggi}  suggests that for proteins with
simple native state topology a suitable one-dimensional
reaction coordinate suffices. In fact, the simple one-dimensional 
Kramers theory can be used to
analyze our results quantitatively. According to Kramers theory the
rate for barrier crossing in the moderate to high viscosity regime is
given by \cite{Hanggi,Chandra} 
\begin{equation}
k_{KR} = \frac{\omega_{A}}{2\pi\omega_{B}}\Bigl(\sqrt{\frac{\zeta^{2}}{4} +
\omega_{B}^{2}} - \frac{\zeta }{2}\Bigr)\exp(-\frac{\Delta F}{T}),
\label{KF}
\end{equation}
where \(\Delta F\) is the typical barrier height that the protein
overcomes en route to the native state, \(\omega _{A}\) and \(\omega _{B}\) 
are the frequencies at the minimum and saddle (transition
state \cite{comment2}) 
points of a {\em suitable undetermined} one-dimensional reaction
coordinate describing the folding process. One of us has argued that
the typical free energy barrier in the folding process scales as
\(\Delta F \simeq \sqrt{M} T_{s}\) \cite{Thirum95,comment1,comment3}. 
If this result is used then there
are two parameters in \(k_{KR}\), namely \(\omega _{A}/\omega _{B}\)
and \(\omega _{B}\), that can be used to fit the simulation
results. The solid lines in Fig. (2) show the results of such a
fit. It is clear that Eq. (\ref{KF}) fits the data quantitatively.  
The numerical values of \(\omega _{A}\) and \(\omega _{B}\) (see
caption to Fig. (2)) for the
four sequences (data for sequence I not shown) suggest that the barriers for
slow folders are, in general, flatter than for fast folders.

The quantitative description of the simulation results by the Kramers
theory shows that (a) the assumption of a one-dimensional reaction
coordinate for folding of proteins with simple native state  topology
is appropriate \cite{Socci96} (Furthermore, it appears that in the
energy diffusion regime an appropriate folding reaction coordinate 
couples rather weakly to the other degrees of freedom. This is
supported by the notion that in nearly all the sequences examined once
a critical number of native contacts is established a rapid
acquisition of the native state takes place);  
(b) The typical barriers in the  folding process scales
sublinearly with \(M\), the number of amino acids, and is adequately
given by \(\sqrt{M}T_{s}\); (c) The transition state estimate for
folding rate in the viscosity regime of experimental interest is at
least two orders of magnitude less than the actual rate. 

In our earlier studies we have shown that due to topological
frustration the refolding of proteins follows the kinetic partitioning
mechanism (KPM) \cite{Thirum95,ThirumWood,Veit}. 
It is of interest to compute  the
partition factor \(\Phi \), which gives the yield of native molecules
that arrive rapidly without being trapped in any intermediate, as a
function of \(\zeta \). Using  the distribution of first passage times
\(\Phi \) can be easily obtained \cite{Veit}. The partition factor \(\Phi \)
shows no  significant variation for the fast folding sequences G and
I (data not shown). 
The dependence of \(\Phi \) on \(\zeta \) for sequence E is
displayed in  Fig. (3). Similar results are obtained for the \(\alpha
\)-helix.  This figure shows that, as suggested earlier
\cite{Thirum95,ThirumWood}, the generic feature of KPM, namely, that
for foldable sequences a fraction of molecules \(\Phi \) (determined
by \(\sigma \)) reaches the native conformation rapidly,  remains
valid for all values of \(\zeta \). Just as the rate of folding itself
\(\Phi \) also shows a non-monotonic behavior. Although there is no
systematic trend in the variation of \(\Phi \) with
\(\zeta \), the sequence E  tends to behave as a fast folder  (\(\Phi
\gtrsim 0.9\)) at the higher and lower values of \(\zeta \).

The large variation of \(\Phi \) with \(\zeta \) for moderate and slow
folders 
suggests that pathways by which the polypeptide chain reaches the
native state can be altered significantly by changing \(\eta \). 
In order to probe
this we generated two hundred distinct initial conditions at \(\zeta =
5.0\) for sequence E. At this value of \(\zeta \) we find \(\Phi \approx
0.8\), and accordingly we determined that there are 
44 trajectories that get trapped
in the CBAs. Using exactly the same initial conditions we altered
\(\zeta \) to 0.16 and discovered that out of the 44 slow folding 
trajectories 20 of them became fast folding indicating the dramatic
change in pathways with the alteration of viscosity. Similar results
were obtained for the 156 fast folding trajectories at the higher
\(\zeta \). Note that \(\Phi \) is a dynamic quantity
and the conservation of the number of denaturated molecules only
requires that the sum of amplitudes of fast and all the 
slow folding phases be
constant. An important experimental consequence of this result 
 is that the folding
scenarios can also be dramatically changed by varying \(\zeta \), so
that a sequence that appears to be the fast folder at one value of
\(\zeta \) may be a moderate folder at a different viscosity.

The results in Fig. (2) may be used to obtain time scales of folding
of small proteins.  The friction coefficient 
\(\zeta \) corresponding to water (\(\eta
=0.01\, Poise\) at \(T=25^{\circ} C\)) is roughly \(6\pi\eta a\) and
this corresponds to \(\zeta \approx 50\) in reduced units with
\(a\approx 5\, \AA\). In this range of \(\zeta \) values the
inertial terms are irrelevant and the natural measure of time is
\(\tau _{H} = \zeta a^2/k_{B}T_{s}\), which for water turns out to
be about \(3\, ns\) \cite{Veit}. 
Using this mapping we find that the time constant
for the formation of \(\beta \)-sheet ranges from \((0.03 - 0.1)\,ms \)
depending on the value of \(\sigma \). A
similar calculation for sequence H shows that the time scale for the
formation of a short \(\alpha \)-helix, containing about four turns, is
about \(10\,\mu s\). It should be noted that the \(\alpha \)-helix in
our study is
not well optimized, since  \(\sigma =0.75\) is relatively large. 
Well optimized sequences have \(\sigma \) values in the range of 0.3
or less \cite{Veit}. If we assume that folding time scales roughly as \(\sigma
^{3}\) \cite{Thirum95} then we predict that an optimized helical
sequence with four turns would fold in almost \(500\,ns\) or 
\(0.5\,\mu s\).

Our theoretical predictions can be verified 
by experiments on folding kinetics in the viscosity regime \(\eta
_{water} \le \eta  \le 10\,\eta _{water}\). 
The  prediction that yield of the fast folding process (namely,
\(\Phi \)) for moderate and slow folders can be drastically 
altered  by changing viscosity is amenable to experimental tests.

\acknowledgments
This work was supported in part by a grant from the National Science
Foundation (CHE96-29845).

\begin{figure}

Fig. 1. The conformations of the native state for sequences I (\(\beta
\)-sheet, left panel) and H (\(\alpha \)-helix, right panel).  The
sequences consist of hydrophobic \(B \) residues (shown in blue),
hydrophilic \(L \) residues (shown in red), and neutral \(N\) residues
(shown in grey).  In the \(\beta \)-sheet neutral residues are  near
the loop region, where the dihedral angles can adopt  \(g^{+}\)
(\(\approx 60^{\circ }\)), \(t\)(\(\approx 180^{\circ }\)), or
\(g^{-}\)(\(\approx -60^{\circ }\))  positions, where \(g^{+}\),
\(t\), and \(g^{-}\) are the three minima in the dihedral angle
potential \cite{Veit}.  
The dihedral angles  in the \(\beta \)-strands are in the
\(t\) conformation. The \(\beta \)-sheet  is stabilized by the
attractive hydrophobic \(B-B\) interactions.  All the dihedral angles
in helical structure are in  \(g^{+}\) positions. The number of beads
(residues) in one turn of the helix is 3.9. These structural features
are in accord with those seen in real proteins. This figure is located
at http://www.glue.umd.edu/\~{}klimov/seq\_I\_H.html.

Fig. 2. The ratio of the folding rate to that of the transition state
value as a function of \(\zeta \). The top, middle, and the bottom
panels correspond to sequences G, E, and H, respectively. Sequences G
and E have a \(\beta \)-sheet as a native state and the native
conformation for sequence H is an \(\alpha \)-helix. The solid lines
are the fit using the Kramers expression for barrier crossing
(cf. Eq. (\ref{KF})). The free energy barrier in Eq. (\ref{KF}) is
taken to be \(\sqrt{M}T_{s}\) so that \(\omega _{A}\) and \(\omega
_{B}\) are used as fitting parameters. From the best fit the
transition state estimate of the folding rate 
is calculated using \(k_{TST}=\omega_{A}
exp(-\sqrt{M})/2\pi\). For sequences G and H the fit is done using
nine data points with \(\zeta \ge 0.16\), while for sequence E ten
data points are used with \(\zeta \ge 0.05\). The most accurate least
squares fit for sequences G, E, and H give [1.86\,(0.01), 2.47\,(0.01)],
[1.95\,(0.08), 1.43\,(0.03)], and [3.78\,(0.02), 0.75\,(0.002)],
respectively. The set of the numbers in the square brackets
corresponds to \(\omega _{A}\) and \(\omega _{B}\) and the numbers in
the parenthesis are error estimates. 
Viscosity for water gives \(\zeta = 50\).

Fig. 3. The  partition factor \(\Phi \), which gives
the fraction of fast folding trajectories,
as a function of \(\zeta \) for sequence E (a moderate folder). 
Folding pathways strongly
depend on \(\zeta \) leading to dramatic variation in \(\Phi \). 
For example,
at \(\zeta = 500\) this sequence may be classified as fast folder
(\(\Phi > 0.9\)), while at all other \(\zeta \) it is a moderate
folder. There is no dependence of \(\Phi \) on \(\zeta \) for fast
folding sequences (data not shown).

\end{figure}

\newpage

\begin{center}
\begin{minipage}{35cm}
\[
\psfig{figure=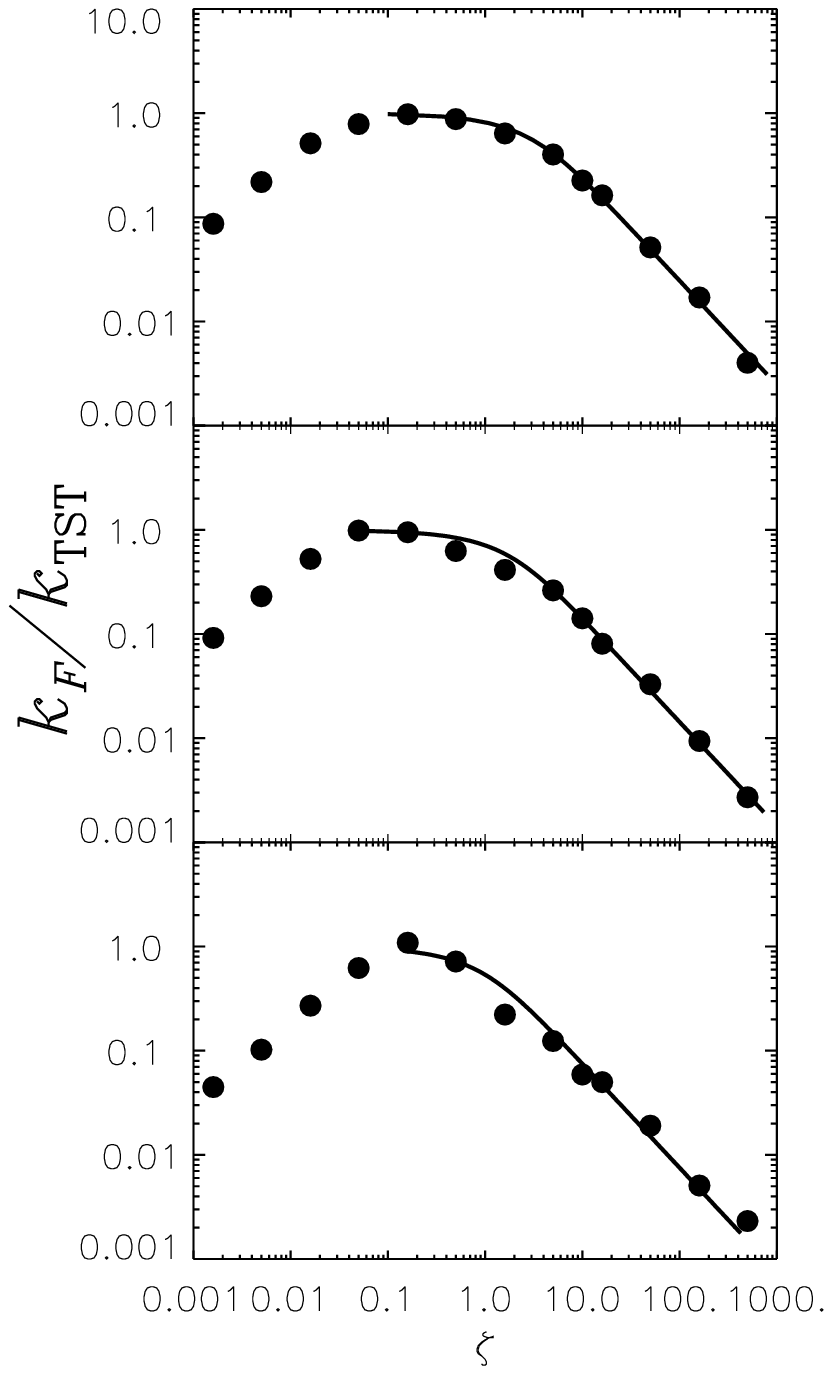,height=20cm,width=35cm}
\]
\end{minipage}
\end{center}

\vspace{1cm}
\begin{center}
{\bf Fig. 2}
\end{center}

\newpage

\begin{center}
\begin{minipage}{18cm}
\[
\psfig{figure=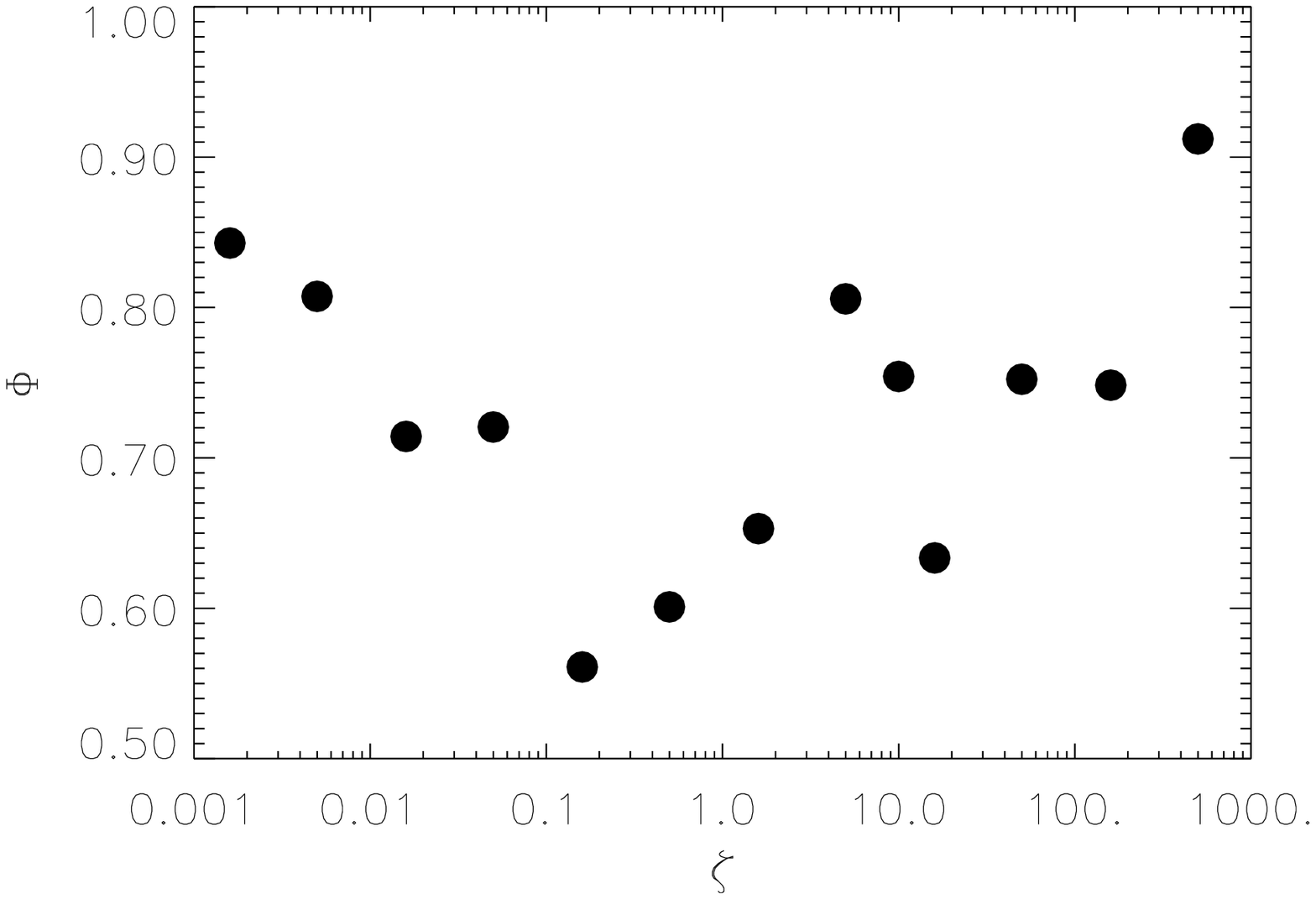,height=14cm,width=18cm}
\]
\end{minipage}

{\bf Fig. 3}
\end{center}


\begin{references}
\bibitem{comment1} It was argued in \cite{Thirum95} that the free
energy barrier scaling as \(\sqrt{M}\) holds good only for
foldable sequences. This result is expected to be valid only over 
a temperature range in which folding rate is optimal. At
very low temperatures the free energy barrier, even for foldable
sequences, could scale as \(M\).
\bibitem{comment3} We have also used \(\Delta F\) as a fitting
parameter and find that the resulting three parameter fit gives a
value very close to the theoretical prediction \(\Delta F \simeq
\sqrt{M}T_{s}\). Higher powers in \(M\) do not fit the data well.
\bibitem{comment2} In proteins there are multiple transition states so
that \(\omega_{B}\) (and  \(\omega_{A}\) as well) will be functions of
\(\chi \). For simplicity these effects are not
considered.
\end{references}
\end{document}